# SPIKING NEURONS WITH ASNN BASED-METHODS FOR THE NEURAL BLOCK CIPHER

Saleh Ali K. Al-Omari[1], Putra Sumari[2]

[1,2] School of Computer Sciences,University Sains Malaysia, 11800 Penang, Malaysia
Salehalomari2005@yahoo.com , putras@cs.usm.my

## ABSTRACT

*Problem statement: This paper examines Artificial Spiking Neural Network (ASNN) which inter-connects group of artificial neurons that uses a mathematical model with the aid of block cipher. The aim of undertaken this research is to come up with a block cipher where by the keys are randomly generated by ASNN which can then have any variable block length. This will show the private key is kept and do not have to be exchange to the other side of the communication channel so it present a more secure procedure of key scheduling. The process enables for a faster change in encryption keys and a network level encryption to be implemented at a high speed without the headache of factorization. Approach: The block cipher is converted in public cryptosystem and had a low level of vulnerability to attack from brute, and moreover can able to defend against linear attacks since the Artificial Neural Networks (ANN) architecture convey non-linearity to the encryption/decryption procedures. Result: In this paper is present to use the Spiking Neural Networks (SNNs) with spiking neurons as its basic unit. The timing for the SNNs is considered and the output is encoded in 1's and 0's depending on the occurrence or not occurrence of spikes as well as the spiking neural networks use a sign function as activation function, and present the weights and the filter coefficients to be adjust, having more degrees of freedom than the classical neural networks. Conclusion: In conclusion therefore, encryption algorithm can be deployed in communication and security applications where data transfers are most crucial. So this paper, the neural block cipher proposed where the keys are generated by the SNN and the seed is considered the public key which generates the both keys on both sides In future therefore a new research will be conducted on the Spiking Neural Network (SNN) impacts on communication.*

## 1.    INTRODUCTION

Over the last one and a half decades, the networking is the driving force among various forms in communications where the progression in the networking had been growing at a phenomenal rate. To ensure growth in networking, the urgent high bandwidth data transfer is important to ensure the continuous growth in the networking. This we can see in today's application in every financial intuition which all done in online for example as video surveillance.

Virtual private network (VPN) is a private communications network which it often used by companies and organization for sending data (voice, video, or group of these media) across secured and encrypted private. The use of IPsec (IP security) for authenticating or encrypting each IP  packet in a data stream channels between two points have brought in the question adequate security mechanisms to ensure the data are transferred safely and securely with better performance levels in order to protect and safeguard sensitive information from outside intruders. In this paper, we also discussed about classes which is fragment into two classes of key-based encryption algorithms, symmetric and asymmetric algorithms. The differences between these two classes are the symmetric algorithms which it used the same key for encryption and decryption, but other type whereas asymmetric algorithms use a different key for encryption and decryption. The decryption key cannot be derived from the encryption key, also the symmetric algorithms fragments into two type, the first one is the stream ciphers and the second one is block ciphers. Block Cipher algorithms are a type of Symmetric Key Encryption. Symmetric Key encryption [19] is a key-based encryption in which the same key that is used to encrypt sensitive data is used to decrypt the sensitive data [18]. Currently, the symmetric algorithms become so familiar and very important to use it [1, 2]

This paper is organized as follows: Section 2 related work describes; section 3 describes the proposed solution. Section 4 the overview of neural cryptosystem in, section 5 describes the neural cryptosystem, and Conclusion in section 6.

## 2. RELATED WORK

Nowadays, the symmetric algorithms fragment into two type the first one is the Stream Ciphers which encrypt a single bit of the plaint text at the same time. But the in the block ciphers take the number of bit and make encrypt them as a single unit. Asymmetric ciphers in the same can call it as public-key algorithms or some time they call it as generally public-key cryptography permit the encryption Key to be public, allowing anyone to encrypt with the key, whereas only the proper recipient (who knows the decryption key) can decrypt the message. The encryption key is also called the public key and the decryption key the private key or secret key.

Modern cryptographic algorithms are no longer pencil and paper ciphers. Strong cryptographic algorithms are designed to be executed by computers or specialized hardware devices. In most applications, cryptography is done with computer software.

Usually, the Symmetric Algorithms are much better and faster to execute on a computer than asymmetric ones. With the practice they are often used together, so the public key is used to encrypt a randomly generated encryption key, and the random key is used to encrypt the actual message using a symmetric algorithm. This is sometimes called hybrid encryption. And we discuss about of many good cryptographic algorithms are widely and publicly available from any major bookstore, patent office, or may be over the Internet. The Data encryption standard has been the major encryption standard for the may several years before. Current trends have drastically impact on distributed spiking neuron network (SNN)[3,4,5,6,7] for handling low level visual perception for extract salient locations in robot camera images (RCI)[1].also they shown a new method which reduce the computational load of the all system, stemming from their chose of the a architecture.

As well as they also explained   about the modelling for the Post Synaptic Potential (PSP), that allows to quickly computing the contribution of a sum of incoming spikes to a neuron's membrane potential. The interests of this saliency extraction method, which differs from classical image processing, are also exposed. Also in the data encryption standard, there are a lot of block ciphers has been proposed as an Advanced (ES) Of which MARS [8], RC6 [9], Serpent [10] and the important Encryption Standard is the Twofish [11] were selected as finalist candidate algorithms. Key Scheduling is the process in which round keys are generated subjecting to a main key processed by a nonlinear function and producing expanded keys and the Key Scheduling is used it in a lot of algorithms [12]. A poor key generation is easily vulnerable to differential and linear cryptanalysis [13]. Giving the importance of the process of generating the key, some research has exploited the use of classical artificial neural networks [14, 15, 16, and 17].

In another related close to my work was a problem arises when a big number of clients or people want to exchange secret messages using a cipher and a secret key. When there's no anyone else than the two involved parties should be able to decrypt the exchanged message, pair wise secret keys want to be distributed among the participants. The benefit of the public key in this case is that they allow to transmission of secret messages without the need to exchange a secret key, and this kind of a general framework for civilizing the security of the cryptosystem based on two important things the first one the symmetric block cipher and the second the key management and also this one based on the asymmetric public key. As well as the main idea is use the benefit of characteristics of Symmetric and Asymmetric.

## 3. PROPOSED SOLUTION

To achieve the goal a variety of security algorithms mechanisms have been designed and these mechanism are collectively known as ciphers, which are used for encrypting and decryption and are difficult to break. In recent years modern ciphers apply key(s) for controlling encryption and decryption and messages are decrypted by the receiver only if the key matches the encryption key. The two categories of key are symmetric key also called secret key and asymmetric also known as public key. Symmetric key apply the same key for both encryption and decryption, in the case of asymmetric key, it applies two set of key for encryption and decryption respectively.

Symmetric key is implemented in two ways either as a block cipher or stream cipher. Block cipher transforms a fixed-length block of plaintext say a fixed size of 64 data into a block of ciphertext (encrypted text) data of the same length. Asymmetric on the other hand allow encryption key of data to be made public for anyone intending to encrypt while only the recipient had access to the private key for decryption.

Research on cryptographic mechanism had proved that symmetric algorithm is quicker to execute on a computer than asymmetric algorithm because of the use of one key for both operations. However in practice both keys (algorithms) are used together to encrypted and decrypt.

Data Encryption Standard (DES) which uses 64 bits block and key was the main encryption standard for applications developments when it first began in 1977. However this algorithm is susceptible to brute force attacks so more block ciphers had been developed to enhance security of data namely Advanced Encryption Standard (AES), MARS, RC6, Serpent, Twofish. Among these algorithm AES had been chosen as the standard algorithm for internet and hardware security because the algorithm is user friendly to computer computational power. But one thing of great interest is that these algorithms apply Key schedule which is an algorithm that, given the key, calculates the sub-keys for each round. Whenever a key generation is weak it becomes easy for differential and linear cryptanalysis that's why research had already been done on the application of classical Artificial Neural Networks (ANN) to address these security concerns. However, Spiking Neural Networks (SNN) which uses simulated neurons for each basic unit of operations and timing of outputs are encoded in 1's and 0's, with 1 meaning it is occurring and 0 being it is not happening. SNN works by using a sign function as activation function and then project the weights and filter co-efficient to be adjusted so allowing for additional independence than ANN.

To rectify the private key issues and applied it in SNN, a prototype symmetric cipher block which had a block length of 8 bits but which in practice can be expanded to either 64 bits, 128 bits or to any variable desired length. The new algorithm, the keys will do not have to be exchanged among parties so key scheduling is very strong, for the simply reason being that the keys are on both sides of the communication channels so are generated by the SNN at each end.

Simple-Data Encryption Standard (S-DES), the block cipher chosen for the research as it is the short version of DES and recognized for most internet security purposes, and moreover the algorithm makes it possible for quicker changes in encryption keys and network level encryption at a much higher speed without the concern of factorizations. With the procedures outline block cipher undergoes changes too in the public cryptosystems.

## 4.   OVERVIEW OF NEURAL CRYPTOSYSTEM

S-DES as shown below (Fig 1) is block cipher and the same keys used for encryption and decryption.  Here, 8 bits block of plaintext for encryption, 10-bit key as input and lastly (8 bit) block of ciphertext as output. And this diagram showed how it's work.

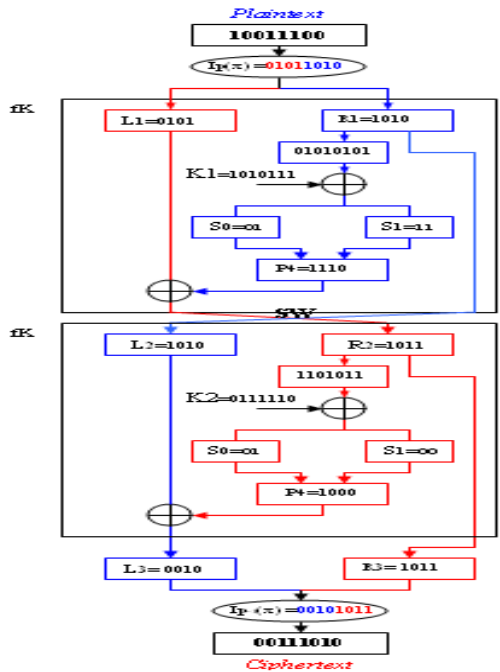

Figure 1. the Simple Data Encryption Standard Block Diagram.

In the case of decryption 8 bits block of ciphertext, 10 bits key as input an (8 bits block of plaintext) as output shown in and we will discuss the result results for the input and output down (Fig 2).

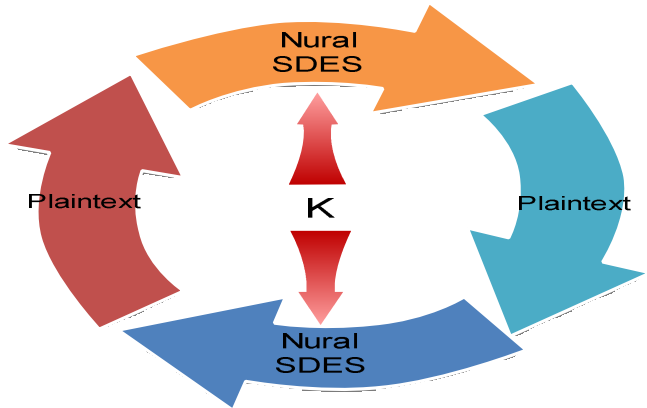

Figure 2. the Neural Cryptosystem

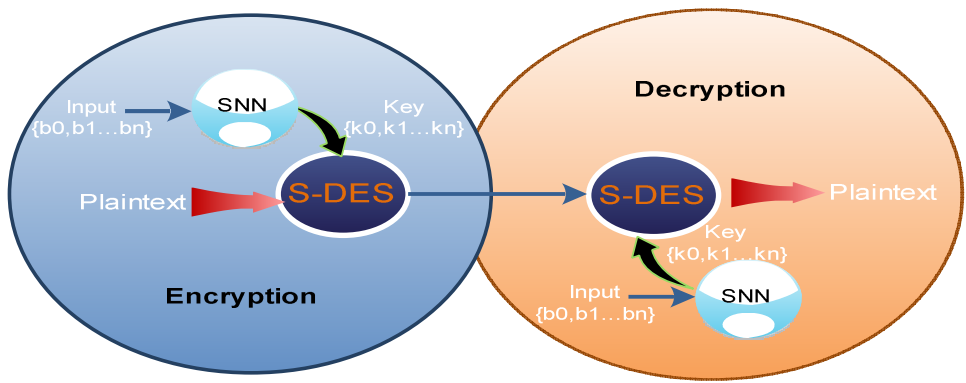

Figure 3. The encryption and the decryption phases

## 4.1 The Algorithm

Up is the algorithm, which includes five functions:

**1.** The Initial permutation (IP);

X = [11011110];

IP (263144857)

IP(x) = [01000010]

**2.** The function, fx, includes both permutation and substitution operations with a sub-keys dependant. The sub-keys as said earlier are generated on both sides on the channel by SNN. Also both sender and receiver share the network by means of secure channel and then receive the seed of the key as public key. The output generated from the spiking neurons are pluses and represents 0's and 1's so with the output display, binary data are formed.

The two procedures use to generate a key are;

**(a)** A 10 bit word passes through a neuron and generates 2 8 bit sub-keys in only one output.

**(b)** Lastly, more than one neuron are used where a 16 bit word passes through two neurons and each one generates a 8 bit sub-keys 3 and since the neurons is non-linear it makes very hard to derive the sub-keys from the public input key. See below

fk (L,R) = (L ⊕ F(R,SK), R)

Where F is a map from 8- bit strings to 4 bit strings and SK is the sub-key K1 or K2.

**3.** An easy permutation function which can switch the two halves of the data.

**4.** The function of fk

**5.** Permutation function which is the inverse of the initial one shown in function1 above.

IP-1 (41347286)

IP-1 ([01000010]) = [10000100]

To finish explaining the fk, the mapping of F function is indicated.

## 4.2 Expansion and Permutation Discussed

**1.** Expansion and permutation operation:

E/P = (41233341)

E/P (0110) = [00100100]

**2.** XOR of output 1 with the first key, K1

**3.** S-boxes are table functions with 1st and 4th bits detailed row, and 3rd bits specify detailed a column which is generating 2 bits output. The first 4 bits are fed to S-box S0; the second 4 bits are fed to S1. These S-boxes, S0, S1 is seen in (Fig 4).

Figure 4. S-boxes table

**4.** The output of 3 permutated as follows;

   P4 = (2431)

   P4 ([1100]) = [1001]

The output of 4, is the output of the F function and is XORed with the left side of the first permutation output. The switch function substituted the left with right so that the second instance of fk operates on a different four bits using K2. It is repeated in the second block and in the end there is no switch and the inverse permutation.

## 5.   NEURAL CRYPTOSYSTEM

The creation of a cryptosystem is founded on the S-DES algorithm in which the key and sub-keys are generated by a spiking neural network.

The first part is to generate the keys for the system to encrypt and decrypt the plaintext. The data that is going through the channel is the ciphertext and the input goes to the networks.

The explanation of the algorithm is as follows:

**1.** Computation of Keys;

   **(a)** Generate seed for input key. This seed generated is the public key and is the input for the network, so the two rounds of seeds generated are X1 and X2

   **(b)** X1 and X2 are inputs into the SNN which had constant boundary

   **(c)** Neurons creates 2 outputs with 8 bits each, and there are the sub-keys to the encryptions algorithm K1 and K2

**2.** Encryption phase:

   **(a)** IP function is applied to the plaintext.

   **(b)** Divide result of (a) into two parts, L1 and R1.

   **(c)** F function applied in R1.

**3.** Expansion, permutation and XOR by means of K1

   E/P = (41232341)

   E/P R1 = E/P ( R1 )

   R'1 = E/P (R1) $\oplus$ K1

**4.** (4 bits) first originated from the R'1 and divided into two parts before going through the S-boxes and generates two output of two bit each, these two are joined together to become one 4 bit strings. See diagram (Fig4).

Moreover,

S0 = S-box S0 ( left  ( R'1 ) )

S1 = S – box S1 ( right ( R'1 ) )

S = [S0 S1]

**5.** An additional permutation function promulgated here;

P4 = (2431)

P4S = P4 (S)

6. P4S is XORed with L1.

**(d)** Switch function exchange left and right, L2 = R1 and R2 = P4S $\oplus$ L1

**(e)** Duplicate the same process from (c) until (d) using the K2. It is not necessary to apply the switch function in the last block.

**(f)** Apply the inverse permutation function to L3 and R3.

IP - 1 ( 41357286 )

Ciphertext = IP-1 ( [ L3 R3 ] )

**7.** Drive the ciphertext and the public key feed to the network into the communication channel

**8.** Decryption phase: The network configuration is established at first and is shared between sender and receiver, so that the receiver computes the keys feeding the input sent through the

Network, The decryption algorithm is similar use to encrypt but the keys are used in the reverse order as seen below;

**Plaintext = DSDES ( K1 ,  K2 ) ( Ciphertext )**

SNN neurons go round with choices of free parameters which are utilized and the pseudo-random key generator and the decided network parameters. When the output changes, the network configuration changes as well and the input feed.

**THE PROPOSAL OUTPUT IS SHOWN BELOW:**

Proposal one:

Plaintext = Saleh al-omari
Ciphertext = Úí~Ëó$\mathrm{G}$ĴÖűc¸6ýò
Key 1         = 00000110
Key 2         = 01111100

Decrypted = Saleh al-omari

Filter constant = 0.001 and 0.0001, weight = 0.5 also time constant = 0.001

The A simulation specifying the constants of a neuron: filter constants equal to 0.01 and 0.0001, weight equal to 0.5, T (time constant) equal to 0.001.

Proposal Tow:

Plaintext = Saleh al-omari
Ciphertext = ý2Ë¸c$\mathrm{G}$Ĵ~űÓ6þHŒ
Key 1         = 10101010
Key 2         = 00110010

Decrypted = Saleh al-omari
Filter constant = 0.01 and 0.0001, weight = 0.2 also time constant = 0.001.

The simulation specifying the constants of a neuron: filter constants equal to 0.1 and 0.0001, weight equal to 0.2, T (time constant) equal to 0.001.

To depict how the simulation works, the word "*Saleh al-omari*" was encrypted. ciphertext, the keys and the SNN configuration.

For the second simulation the same plaintext was encrypted, however the weight and filter constant changes because of there was changes in the sub-keys and the cipher text as shown in (Proposal 1, 2).

The simulation of SNN has the capability to generate sub-keys with non-linear function that comes from the proposed model. S-DES is a public cryptosystem, and the input of SNN is the public key and the private key is calculated by the feeding the public through the SNN network. The 8 bit key applied here was to speed up the simulation and so by increasing the number of keys bit to 64, 128 or any required length so as to make it stronger.

# 6.    CONCLUSION

The block cipher is converted in public cryptosystem and had a low level of vulnerability to attack from brute, and moreover can able to defend against linear attacks since the Artificial Neural Networks (ANN) architecture convey non-linearity to the encryption/decryption procedures. In this paper is present to use the Spiking Neural Networks (SNNs) with spiking neurons as its basic unit. The timing for the SNNs is considered and the output is encoded in 1's and 0's depending on the occurrence or not occurrence of spikes as well as the spiking neural networks use a sign function as activation function, and present the weights and the filter coefficients to be adjust, having more degrees of freedom than the classical neural networks. In conclusion therefore, encryption algorithm can be deployed in communication and security applications where data transfers are most crucial. So this paper, the neural block cipher proposed where the keys are generated by the SNN and the seed is considered the public key which generates the both keys on both sides, and in future therefore a new research will be conducted on the SNN impacts on communication.

## ACKNOWLEDGEMENTS

Special thank and recognition go to my advisor, Associate Professor. Dr. Putra Sumari, who guided me through this study, inspired and motivated me. Last but not least, the researchers would like to thank the University Sains Malaysia USM for supporting this study.

**Authors**

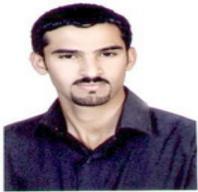

**Saleh Ali K. Al-Omari** Obtained his Bachelor degree in Computer Science from Jerash University, Jordan in 2004-2005 and Master degree in Computer Science from Universiti Sains Malaysia, Penang ,Malaysia in 2007. Currently, He is a PhD candidate at the School of Computer Science, University Sains Malaysia, Penang, His main research interest are in Video on Demand (VoD) over Heterogeneous Mobile Ad Hoc Networks (MANETs).

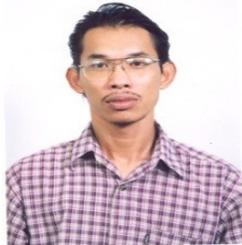

**Putra Sumari** is currently working as Assistant Professor of School of Computer Science, Universiti Sains Malaysia, Penang 2009. Assoc. Prof Dr. Putra received his MSc and PhD in 1997 and 2000 from Liverpool University, England. Currently, he is a lecturer at the School of Computer Science, Universiti Sains Malaysia, Penang. He is the head of the Multimedia Computing Research Group, School of Computer Science, USM. He is Member of ACM and IEEE, Program Committee and reviewer of several International Conference on Information and Communication Technology (ICT), Committee of Malaysian ISO Standard Working Group on Software Engineering Practice, Chairman of Industrial Training Program School of Computer Science USM, and Advisor of Master in Multimedia Education Program, UPSI, and Perak.